\documentclass{article}
\usepackage{amsmath}
\usepackage{amssymb,latexsym}
\usepackage{graphicx}

\def\ve{\varepsilon}
\def\<{\langle}
\def\>{\rangle}
\def\GeV{{\rm\ GeV}}
\def\MeV{{\rm\ MeV}}

\def\para{\parallel}
\def\Re{\mathop{\rm Re}\nolimits}
\def\Im{\mathop{\rm Im}\nolimits}
\def\tg{\mathop{\rm tg}\nolimits}

\def\CM{{\cal M}}

\def\eps{\epsilon}
\def\ka{\varkappa}

\def\vec#1{{\bf #1}}

\begin{document}
\title{On radiative corrections to polarization observables\\
 in electron-proton scattering}
\author{Dmitry~Borisyuk, Alexander~Kobushkin\\[2mm]
\it Bogolyubov Institute for Theoretical Physics,\\
\it 14-B Metrologicheskaya street, Kiev 03680, Ukraine}
\date{}
\maketitle

\abstract{
 We consider radiative corrections to polarization observables
 in elastic electron-proton scattering, in particular,
 for the polarization transfer measurements
 of the proton form factor ratio $\mu G_E/G_M$.
 The corrections are of two types:
 two-photon exchange (TPE) and bremsstrahlung (BS);
 in the present work we pay special attention to the latter.
 Assuming small missing energy or missing mass cut-off, the correction
 can be represented in a model-independent form,
 with both electron and proton radiation taken into account.
 Numerical calculations show that the contribution of the proton radiation
 is not negligible.
 Overall, at high $Q^2$ and energies the total correction to $\mu G_E/G_M$
 grows, but is dominated by TPE.
 At low energies both TPE and BS may be significant;
 the latter amounts to $\sim 0.01$ for some reasonable cut-off choices. 
}

\section{Introduction}

Polarized and unpolarized elastic electron scattering
are important sources of information about nucleon structure,
which in this case reveals itself via electromagnetic
form factors (FFs).
Study of $Q^2$ dependence of the FFs allows, for example,
to determine nucleon size and quark content,
test phenomenological nucleon models, QCD predictions,
and much more.
However, the elastic scattering amplitude and the FFs are simply related
only in the first order in $\alpha$, the electromagnetic interaction constant.
An analysis of experimental data 
necessarily involves the calculation
of higher-order effects, often called radiative corrections.
In connection with the modern development
of polarization transfer experiments \cite{JLab-expts}
it became clear that higher-order corrections can
seriously influence the FF measurements (see, e.g., Refs.~\cite{howWell,ourPheno,BMT}).
Nevertheless, current cituation with the estimates of the
radiative corrections to polarized $ep$ scattering is not quite satisfactory.

There are two distinct types of radiative corrections:
1) The corrections of the second order in $\alpha$ to the elastic amplitude.
An example of such correction is two-photon exchange (TPE),
and
2) The radiation of undetected (soft) photons, known as bremsstrahlung (BS)
or radiative corrections in a narrow sense.
It is well-known that both corrections are, in general, infra-red (IR)
divergent and a finite cross-section value is obtained
only after their summation.
Hence both corrections should be considered in common.
However, this was never done in calculations of 
the radiative corrections to polarization observables in $ep \to ep$.
Thus, in Refs.~\cite{BMT} the correction was attributed entirely to TPE.
On contrary, in Refs.~\cite{Afa1,Afa2} the BS corrections were thoroughly
analyzed, but the contribution of the box diagram (TPE) was ignored
"because its treatment ... requires different methods".

This became possible, since the IR-divergent term is {\it factorizable};
that is, it can be reduced to an overall spin-independent factor
in the cross-section.
Thus polarization observables, which are, in fact, ratios of some polarized
cross-section to the unpolarized one, never suffer from the IR divergence;
both TPE and BS corrections to such observables
appear finite and may be calculated separately.
Nevertheless, an arbitrary omission of one of them
is obviously incorrect.

During last decade, TPE was studied rather thoroughly, and the corresponding amplitude
was calculated by different authors in various approximations and kinematical conditions;
we will not discuss it in detail here, but will just refer to the known results \cite{BMT,ourDisp,ourP33}.

As to the BS, the most recent works on this subject
\cite{Afa1,Afa2} still have some drawbacks.
The main idea of the papers \cite{Afa1,Afa2} was an exact model-independent
calculation of the BS effects.
To achieve this, authors were to consider
only the radiation by the electron, but not by the proton,
since the latter depends on the details of the proton structure.

The neglection of the proton radiation seems well-justified at low momentum
transfer, when the proton remains practically at rest.
However, at typical experimental conditions in JLab \cite{JLab-expts}
the final proton is relativistic, thus the electron and the proton
are on an equal footing and their contributions to the BS
should be of the same order of magnitude.

Even the full cancellation of the IR divergence is impossible without
taking into account proton radiation.
Namely, if the BS amplitude is $\CM = \CM_e + \CM_p$,
where the subscript indicates radiating particle,
then the IR divergence in $\int|\CM_e|^2 d\Gamma$ cancels with that
  of the electron vertex correction,
the IR divergence in $\int|\CM_p|^2 d\Gamma$ ---
  with the proton vertex correction,
and the IR divergence in $\Re \int \CM_e\CM_p^* d\Gamma$ ---
  with the IR divergence of the TPE amplitude.

In the present paper we analyze radiative corrections
to the polarization observables in $ep \to ep$.
We take into account both TPE and BS, and for the latter
include the radiation by the electron as well as by the proton.
Certainly exact analytical and model-independent calculation
of the proton radiation is impossible
(still this is not needed for practical applications).
However, we are able to obtain the result of such sort after
the expansion in powers of photon energy, in the first non-vanishing order.
The case of polarization transfer measurements
of the proton FFs is considered in detail.
Our approach is also applicable to other polarization experiments,
e.g., measurements of beam-target asymmetry.

So as not to go into details of different experiments,
we consider a simple idealized experiment in which
the final proton is detected in a fixed direction,
that is, the angular acceptance of the proton detector is very small.
Both electron and proton energies are measured
to determine missing energy $\Delta E$, and the event is counted
as the elastic one if $\Delta E < r_m$, where $r_m$ is some cut-off.
This is the way the elastic events were selected
in the real experiments \cite{JLab-expts}.
Authors of Refs.~\cite{Afa1,Afa2} use a cut on the missing mass,
which was not applied in Ref.~\cite{JLab-expts}.
This case is also considered in our paper and compared
with the "missing energy" approach.

To reduce inelastic background, one must choose reasonably small $r_m$.
For example, to exclude pion production,
$r_m$ should be restricted by $r_m < m_\pi \approx 140 \MeV$.
Therefore we have a small parameter $r_m$ or, more precisely,
$r_m/M$ (where $M$ is the proton mass).
We calculate the radiative correction
in the first non-vanishing order in $r_m$.
To this order, the low-energy theorem \cite{Low}
allows us to obtain a model-independent result in the sense that
it is expressed solely through on-shell proton FFs and their derivatives.

\section{Bremsstrahlung cross-section}\label{Sec:BS}

The process under consideration is
\begin{equation}
 e(k) + p(p) \to e(k') + p(p') + \gamma(r).
\end{equation}
We will also use the alternative notation $k_a$ for particle momenta,
$k_1 = k$, $k_2 = p$, $k_3 = k'$, $k_4 = p'$.
Throughout the paper space components of 4-vectors
are denoted by boldface, e.g., $\vec k$, $\vec p$, etc.
The electron and proton masses are $m$ and $M$, respectively.
The cross-section is given by
\begin{equation}
 d\sigma_\gamma = \frac{1}{(2\pi)^5} \delta(p\!+\!k\!-\!p'\!-\!k'\!-\!r)
   \frac{|\CM|^2}{4M \eps}
   \frac{d\vec p'}{2E'} \frac{d\vec k'}{2\eps'} \frac{d\vec r}{2|\vec r|},
\end{equation}
where $\CM$ is the scattering amplitude,
$\eps$ is initial electron energy in the lab. frame,
$\eps'$, $E'$ and $|\vec r|$ are final electron, proton and photon energies.
The appropriate summation/averaging over polarizations is implied.
There are 9 variables here ($\vec p'$, $\vec k'$ and $\vec r$),
but due to the $\delta$-function only 5 are independent.
We choose the independent integration variables to be
$\vec r$ and $\vec n = \vec p'/|\vec p'|$.
This is convenient since $\vec r$ is small due to the missing energy cut
$|\vec r| < r_m$ and $\vec n$ is fixed in our kinematics (see Introduction).
All other kinematical quantities will be functions of $\vec r$ and $\vec n$,
thus
\begin{equation}
 k' = k'(\vec n,\vec r), \qquad p' = p'(\vec n,\vec r).
\end{equation}
Putting $\vec r = 0$, we return to purely elastic scattering, for which we denote
\begin{equation}
 k'_0 = k'(\vec n,0), \qquad p'_0 = p'(\vec n,0).
\end{equation}
Below we will make an expansion in powers of $r$.
Requiring $k'^2 = m^2$ and $p'^2 = M^2$ it is easy to find 
\begin{eqnarray}
 p' & = & p'_0 + \delta p' + O(\vec r^2), \label{Dp}
 \\
 k' & = & k'_0 - \delta p' - r + O(\vec r^2), \label{Dk}
\end{eqnarray}
where
\begin{equation}
 \delta p' = 2 (k'_0 r) \frac{M^2 p - (pp'_0) p'_0}{M Q^2 (\eps+M)}.
\end{equation}
The $\delta$-function can be rewritten as
\begin{equation}
 \delta(p\!+\!k\!-\!p'\!-\!k'\!-\!r) = J(\vec n,\vec r)
   \cdot 2E' \delta(\vec p' - \vec p'(\vec n,\vec r))
   \cdot 2\eps' \delta(\vec k' - \vec k'(\vec n,\vec r))
   \cdot d\Omega_{\vec n},      
\end{equation}
where
\begin{equation}
 J = \frac{1}{4M^2} \, \frac{[(pp')^2-M^4]^{3/2}}{(pp')(p'k')-M^2 (pk')}.
\end{equation}
Thus we have
\begin{equation}
 \frac{d\sigma_\gamma}{d\Omega_{\vec n}} = \frac{1}{(2\pi)^2}
   \frac{1}{4M\eps} \int |\CM|^2 \frac{J d\vec r}{(2\pi)^3 2|\vec r|}.
\end{equation}
The amplitude $\CM$ depends, in particular, on $p'$ and $k'$,
which should be understood as functions of
$\vec r$ and $\vec n$, Eqs.~(\ref{Dp},\ref{Dk}).
Now we want to expand the integrand into the series in $r$,
keeping two leading terms.
The first term will be $O(1/|\vec r|)$ and corresponds
to the so-called Mo\&Tsai approximation \cite{MoTsai}.
Thus
\begin{eqnarray}
 \CM = \sqrt{4\pi\alpha} \, \ve_\mu^* \left\{ \CM_1 
   \left( \frac{k_\mu}{kr} - \frac{k'_\mu}{k'r}
      + \frac{Zp'_\mu}{p'r} - \frac{Zp_\mu}{pr} \right)
   + \delta\CM_{\mu\nu} r_\nu + O(|\vec r|) \right\} = \\
   = \sqrt{4\pi\alpha} \left\{ \CM_1 \sum_a z_a \frac{k_a \ve^*}{k_a r}
   + \sum_a \delta\CM_{\mu\nu a}  \frac{r_\nu \ve_\mu^*}{k_a r} + O(|\vec r|) \right\}, \nonumber
\end{eqnarray}
where $\CM_1$ is the elastic scattering amplitude in the Born approximation,
$Z=1$ and is introduced, as usually, to distinguish
electron and proton radiation (for positron-proton scattering we would put $Z=-1$),
$z_1 = 1$, $z_2 = -Z$, $z_3 = -1$, $z_4 = Z$,
$\ve_\mu$ is photon polarization vector,
and $\delta\CM_{\mu\nu a}$ is independent of $r$.
Due to the low-energy theorem \cite{Low}, $\delta\CM_{\mu\nu a}$
can be expressed via $\CM_1$, that is, via on-shell proton FFs
(for more detail see the next section).
The cross-section will be
\begin{eqnarray} \label{xsect-series}
 d\sigma_\gamma \sim \int |\CM|^2 \frac{J d\vec r}{(2\pi)^3 2|\vec r|} =
 - 4\pi\alpha |\CM_1|^2 \sum_{a,b} z_a z_b (k_a k_b)
   \int \frac{J(\vec n,\vec r)}{(k_a r)(k_b r)} \,
   \frac{d\vec r}{(2\pi)^3 2|\vec r|} - \\
 - 4\pi\alpha J(\vec n,0) \cdot 2\Re \CM_1^* \sum_{a,b} \delta\CM_{\mu\nu b} z_a k_{a \mu}
   \int \frac{r_\nu}{(k_a r)(k_b r)} \,
    \frac{d\vec r}{(2\pi)^3 2|\vec r|}
 + O(r_m^2). \nonumber 
\end{eqnarray}
Note that the overall minus sign appears because
the photon polarization sum is $\overline{\ve_\mu\ve^*_\nu} = -g_{\mu\nu}$.
The first term is IR-divergent. It is well-known that the divergence cancels
with the IR divergence in the TPE correction.
Since that term 
has the same spin structure as the Born cross-section,
it influences only the unpolarized cross-section
(which is not of our interest), but not polarization observables%
\footnote{
 Indeed, if
 $\sigma = \sigma_{\rm Born} + \delta_f \sigma_{\rm Born} + \delta\sigma_{nf}$,
 where $\delta_f$ does not depend on particle spins, then
 $\sigma \approx (1+\delta_f)(\sigma_{\rm Born} + \delta\sigma_{nf})$,
 and the overall spin-independent factor $1+\delta_f$ cancels
 then computing any polarization observables.%
}.
By the same reason, we do not need to expand $J(\vec n,\vec r)$
under the first integral in the l.h.s. of Eq.(\ref{xsect-series}).

So we should consider the second term. It is IR-finite,
and the integrals it contains have the form
\begin{equation} \label{phi}
  \phi_{ab\mu} = \int \frac{r_\mu}{(k_a r)(k_b r)} \, 
    \frac{d\vec r}{(2\pi)^3 2|\vec r|} = 
  \frac{r_m}{16\pi^3} \int \frac{\rho_\mu}{(k_a \rho)(k_b \rho)} \, d\Omega_{\vec\rho},
\end{equation}
where $\rho_\mu=r_\mu/|\vec r|$ depends on angular variables only.
The integrals involving electron momenta are divergent at $m\to 0$:
\begin{equation} \label{mdiv}
 \phi_{11},\phi_{33} \sim A + B\ln m^2 + C/m^2,\qquad
 \phi_{12},\phi_{13},\phi_{14},\phi_{32},\phi_{34} \sim A + B\ln m^2
\end{equation}
(of course the $C/m^2$ terms cancel when computing any observable,
but the logarithmic terms persist).
These integrals are written out in Appendix~\ref{App:integrals}.

It is interesting to note that the second term in (\ref{xsect-series})
can be viewed as a contribution coming from TPE with the amplitude
\begin{equation}
 \delta\CM = -4\pi\alpha\sum_{a,b} \delta\CM_{\mu\nu b} z_a k_{a\mu} \phi_{ab\nu}.
\end{equation}
Similarly to the TPE amplitude, the quantity $\delta\CM$ can be expressed
via scalar invariant amplitudes (generalized FFs).
However, now we have to include not 6, but 8 FFs:
\begin{eqnarray} \label{FullAmpl}
 \CM = \CM_1 + \delta\CM & = & - \frac{4\pi\alpha Z}{q^2}
     \, \bar u'\gamma_\mu u \, \bar U' \gamma_\nu U
  \left\{   (F_1+F_2) g_{\mu\nu}
          - F_2 \frac{P_\mu P_\nu}{M^2}
          + F_3 \frac{P_\mu K_\nu}{M^2}
          + F_4 \frac{K_\mu P_\nu}{m^2}
          - F_5 \frac{K_\mu K_\nu}{m^2} \right\} - \nonumber \\
 && - \frac{4\pi\alpha Z}{q^2} \left\{
    F_6 \, \bar u'\gamma_5 u \, \bar U' \gamma_5 U
    + \frac{i F_7 }{m q^2} \bar u'\gamma_5 u \, \bar U' \hat e U
    + \frac{i F_8 }{M q^2} \bar u' \hat e u \, \bar U' \gamma_5 U
   \right\},    
\end{eqnarray}
where $u$, $u'$ and $U$, $U'$ are electron and proton spinors, respectively,
$P = p+p'$, $K = k+k'$, $q = p'-p$ and
$\hat e = 4 \ve^{\mu\nu\sigma\tau} \gamma_\mu P_\nu K_\sigma q_\tau$.

For the genuine TPE the amplitudes $F_{4,5,6}$ vanish
in the zero electron mass limit and $F_{7,8} \equiv 0$,
since these amplitudes violate T-invariance.
On contrary, for the effective TPE describing BS, even if $m \to 0$,
non-zero contributions arise in $F_{1-5}$ and $F_{7,8}$ 
(the amplitude $F_6$ is identically zero in our approximation).

The appearance of the T-violating amplitudes can be easily understood.
In general, T-invariance connects the amplitudes for direct and time-reversed processes.
For the elastic process the initial and final particles coincide,
thus direct and reverse processes are, actually, the same process,
and T-invariance imposes some constraints on its amplitude.
For the BS an additional emitted photon
breaks the symmetry between initial and final states,
so the effective TPE amplitude need not necessarily
be symmetric under time reversal.

Once we have converted BS into effective TPE,
we may calculate both TPE and BS corrections
to any observable through the same formulae
(technically, this may be not the easiest way, but we find it interesting
and feasible with symbolic calculation software).

The explicit expressions for the effective TPE amplitudes $F_i$ are given
in Appendix~\ref{App:Fi}.

The detailed formulae for the cross-section and other observables
in terms of the amplitudes $F_i$ are given in Appendix~\ref{App:M2}.
Here we write down only the correction to $G_E/G_M$
ratio, measured via polarization transfer.
In the Born approximation, we have
\begin{equation} \label{R}
 R \equiv - \frac{S_\perp}{S_\para} \, \frac{\eps+\eps'}{2M} \tg\frac{\theta}{2}
 = \frac{G_E}{G_M},
\end{equation}
where $\theta$ is lab. scattering angle and $S_\perp$ ($S_\para$) is
the transverse (longitudinal) component of final proton polarization.
The correction to this quantity is given by
\begin{eqnarray} \label{deltaR}
 \frac{F_m^2}{1-\frac{q^2}{4M^2}} \delta R = 
 F_2\delta F_m - F_m \left( \delta F_2 - \frac{4M^2}{\nu\!-\!q^2} \delta F_4 \right)
 + \left( \frac{\nu F_2}{4M^2} - \frac{q^2}{\nu} F_e \right)
   \left( \delta F_3 - \frac{4M^2}{\nu\!-\!q^2} \delta F_5 + 2 \delta F_8 \right) + \\
 + \frac{8M^2}{\nu} F_e \delta F_7
 - \frac{2F_1 F_e \delta F_8}{\nu q^2 F_m}(\nu^2\!+\!q^2(4M^2\!-\!q^2)),  \nonumber 
\end{eqnarray}
where $\nu = 4(p+p')(k+k')$ and the prefix $\delta$ indicates contribution of order $\alpha$.
We have defined $F_m = F_1+F_2$ and $F_e = F_1 + \frac{q^2}{4M^2} F_2$;
these are not the same as the elastic FFs $G_M$ and $G_E$,
since the former incorporate radiative corrections.

\section{Bremsstrahlung amplitude in detail}\label{Sec:BSdetail}

\begin{figure}[b]
 \hfil
 \includegraphics[width=0.15\textwidth]{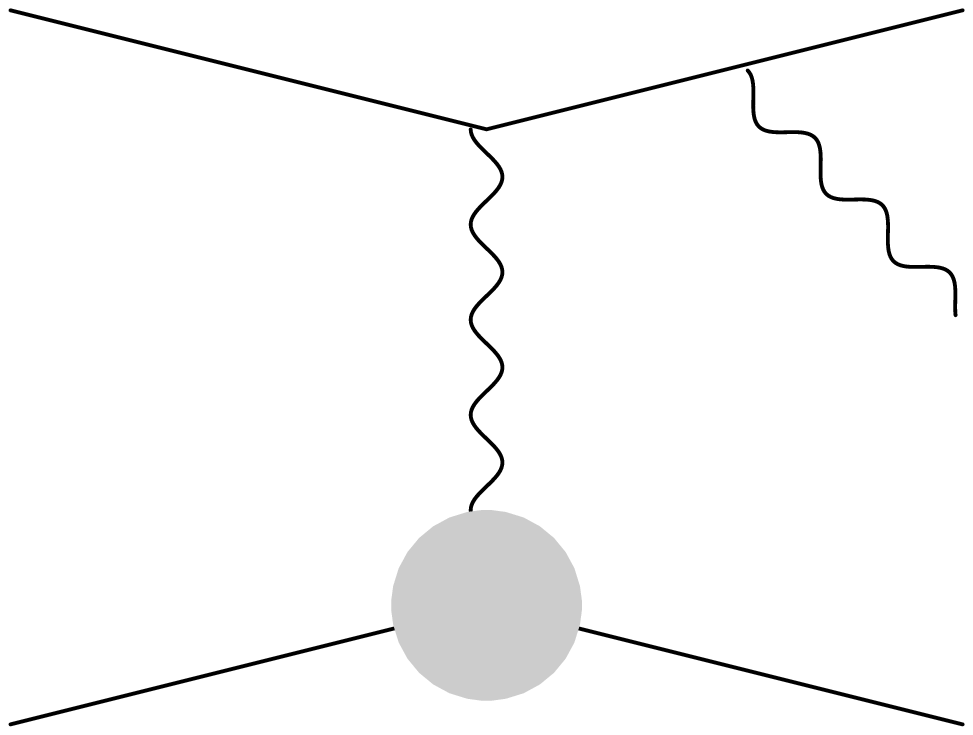}\hfil
 \includegraphics[width=0.15\textwidth]{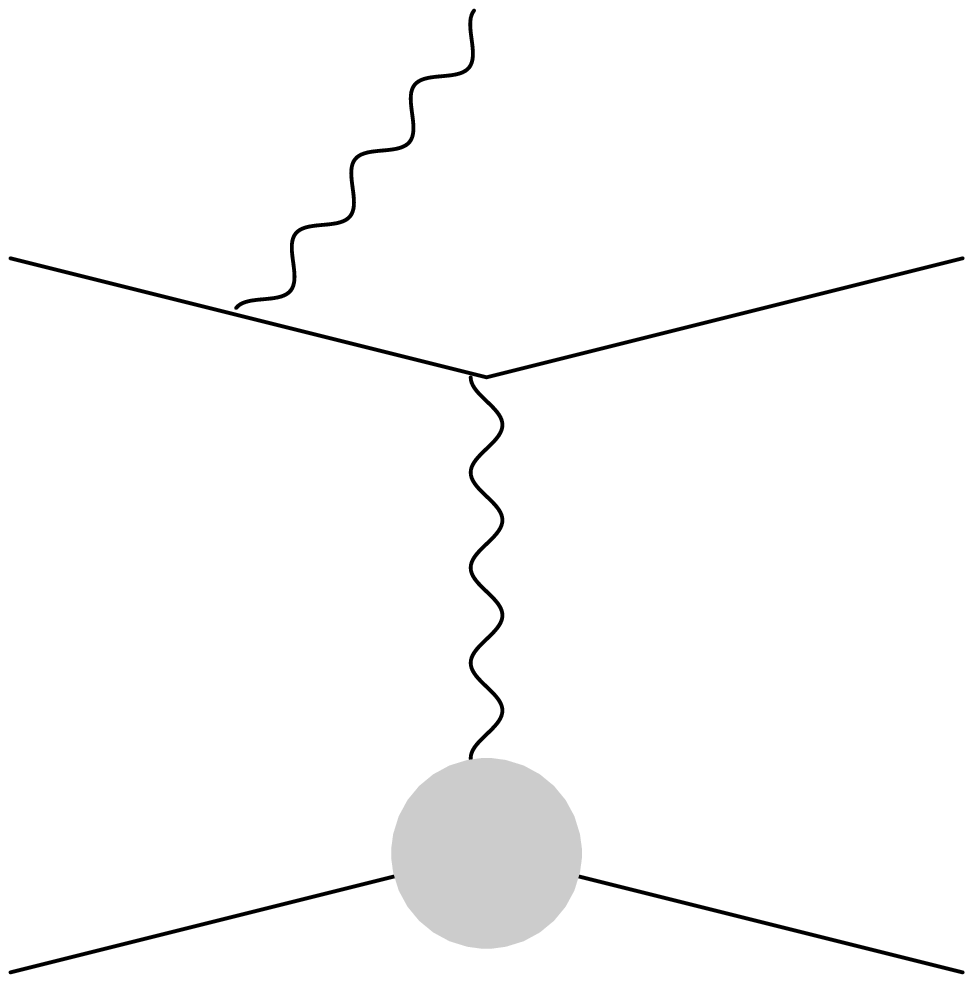}\hfil
 \includegraphics[width=0.15\textwidth]{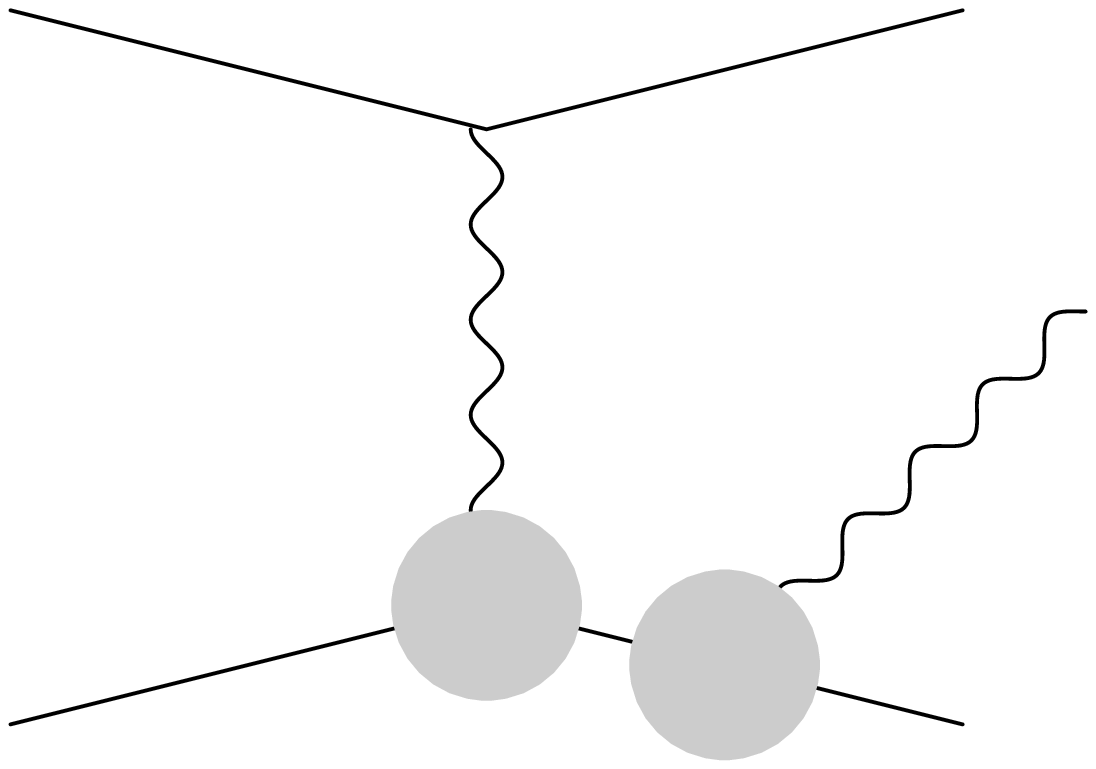}\hfil
 \includegraphics[width=0.15\textwidth]{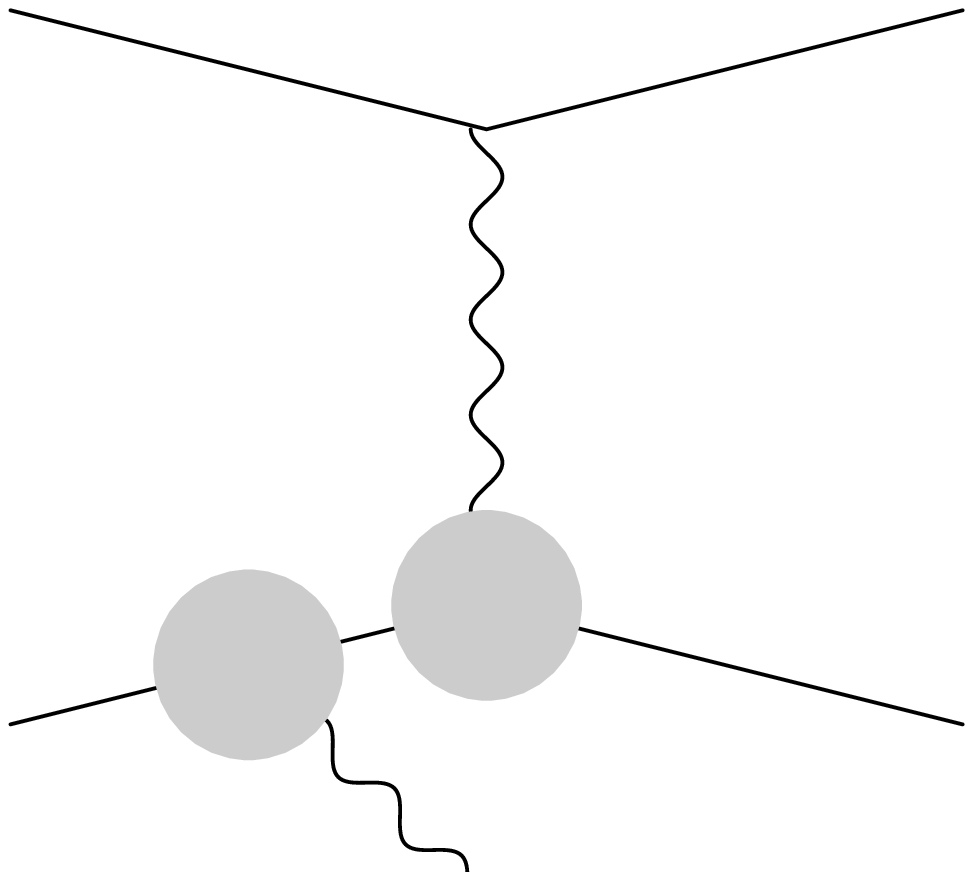}\hfil
\caption{Bremsstrahlung diagrams.}\label{diagr}
\end{figure}
The process amplitude is represented by the four diagrams (Fig.~\ref{diagr}) and equals
\begin{eqnarray} \label{M_g}
 \frac{\CM}{(4\pi\alpha)^{3/2}Z} & = &
  \ve^* \left( \frac{k'}{k'r} - \frac{k}{kr} - \frac{Zp'}{p'r} + \frac{Zp}{pr}
    \right) \bar u' \gamma_\mu u \, \bar U' \Gamma_\mu(q) U + \\
  & + & \left( \frac{1}{2k'r} \bar u' \hat \ve^* \hat r \gamma_\mu u
         + \frac{1}{2kr} \bar u' \gamma_\mu \hat r \hat \ve^* u
    \right) \bar U' \Gamma_\mu(q) U - \nonumber \\
  & - & Z \bar u' \gamma_\mu u \, \bar U'
    \left\{ \frac{\kappa}{4M}
       \left( [\hat \ve^*, \hat r] \frac{\hat p'+M}{2p'r} \Gamma_\mu(q) 
        - \Gamma_\mu(q) \frac{\hat p+M}{2pr} [\hat \ve^*, \hat r] \right)
  + \right. \nonumber \\
  & + & \left. \frac{1}{2p'r} \hat\ve^* \hat r \Gamma_\mu(q)
  + \frac{1}{2pr} \Gamma_\mu(q) \hat r \hat\ve^*
  + \left( \frac{p'\ve^*}{p'r} - \frac{p\ve^*}{pr} \right)
      \left( \Gamma_\mu(q+r) - \Gamma_\mu(q) \right) \right\} U, \nonumber
\end{eqnarray}
where
\begin{equation} \label{Gamma}
 \Gamma_\mu(q) = \gamma_\mu \bar F_1(q^2) -
   \frac{1}{4M} [\gamma_\mu,\hat q] \bar F_2(q^2), \qquad
   \bar F_i(q^2) = F_i(q^2)/q^2, \qquad \kappa = F_2(0).
\end{equation}
The terms without $Z$ correspond to the electron,
the terms containing $Z$ --- to the proton radiation.
In the latter case one needs to take into account off-shell effects.
However to the needed order in $r$ they can be estimated
in a model-independent way using gauge invariance
(this is a sort of so-called low-energy theorem \cite{Low}).
It turns out that these effects are absent in the leading order in $r$.
The argumentation is quite similar to the one given in Refs~\cite{Landau,Low}.

In short, let us assume that FFs in Eqs.~(\ref{M_g},\ref{Gamma})
depend also on the proton virtuality $v$, $F_i(q^2) \to F_i(q^2,v)$.
For the two last diagrams in Fig.~\ref{diagr},
the virtualities are $2p'r$ and $2pr$, respectively.
Thus expanding in powers of $r$ we have
\begin{equation} \label{F-virt}
 F_i(q^2,v) = F_i(q^2,0) + v \frac{\partial F_i}{\partial v}(q^2,0) + O(\vec r^2).
\end{equation}
The first term is the on-shell FF and the second represents sought
off-shell correction. After inserting (\ref{F-virt}) in the full expression
for the ampltude (\ref{M_g}), $v$ from the second term cancels
with the proton propagator, thus the resulting contribution to the amplitude
will be independent of $r$.

On the other hand, gauge invariance requires the amplitude
to vanish upon the substitution $\ve_\mu \to r_\mu$.
This condition allows to determine the off-shell correction unambiguously.
The straightforward calculation shows that the amplitude (\ref{M_g})
is already gauge-invariant, thus ($r$-independent) off-shell
correction should be identically zero.

Now we should carefully expand (\ref{M_g}) in powers of $r$,
following the pattern of Eq.~(\ref{xsect-series}).
The first term in Eq.~(\ref{M_g}) seems to be proportional to the elastic Born amplitude,
but this is not the case: the momenta of final particles $p'$ and $k'$
are not equal to the "elastic" ones $p'_0$ and $k'_0$.
To obtain the correct amplitude expansion, we use the formula
\begin{equation} \label{Delta-u}
 u(p+\delta p, S+\delta S) - u(p,S) \approx \frac{1}{2M}
  \left[ \widehat {\delta p}
    - \gamma_5( M \widehat {\delta S} - p \delta S)
  \right] u(p,S),
\end{equation}
where $p$ is particle momentum, $S$ is its spin 4-vector.

In all cases of our interest it is possible
to use the above equation with $\delta S = 0$.
Indeed, if we consider an experiment with polarized target
(beam-target asymmetry), then the polarization
of final particles is not measured and $\delta S = 0$.
In the polarization-transfer experiment two quantities are measured:
longitudinal and transverse polarizations of the final proton.
Thus we should first insert $p'$ in place of $p$ in Eq. (\ref{Delta-u}).
Since in our calculations the direction of final proton momentum
is fixed and only its magnitude can vary, we easily have
$\delta S = 0$ for the measurement of transverse polarization
and $\delta S = -\frac{S \delta p'}{M^2} p'$ for the longitudinal one.
But the latter expression still yields zero contribution
to the r.h.s. of (\ref{Delta-u}).
So we have
\begin{eqnarray}
 \bar u'\gamma_\mu u \, \bar U' \Gamma_\mu(q) U 
 & = & \bar u'_0 \gamma_\mu u \, \bar U'_0 \Gamma_\mu(q_0) U - \nonumber \\
 & - & \frac{1}{2m} \bar u'_0 (\widehat{\delta p'} + \hat r) \gamma_\mu u
          \, \bar U'_0 \Gamma_\mu(q_0) U + \nonumber \\
 & + & \frac{1}{2M} \bar u'_0 \gamma_\mu u
          \, \bar U'_0 \widehat{\delta p'} \Gamma_\mu(q_0) U + \nonumber \\
 & + & \delta p'_\nu \bar u'_0 \gamma_\mu u
          \, \bar U'_0 \frac{\partial\Gamma_\mu}{\partial q_\nu}(q_0) U + O(\vec r^2),
\end{eqnarray}
where $q_0 = p_0'-p$ and $\delta p'$ is from Eq.~(\ref{Dp}).
There is no need to expand the bracket
$\frac{k'}{k'r} - \frac{k}{kr} - \frac{Zp'}{p'r} + \frac{Zp}{pr}$
in Eq.~(\ref{M_g}), since the resulting expression
will be anyway proportional to the Born amplitude and does not influence
polarization observables. The Born amplitude is (remember that the photon propagator is included into $\Gamma_\mu$)
\begin{equation}
 \CM_1 = -4\pi\alpha Z \, \bar u'_0 \gamma_\mu u \, \bar U'_0 \Gamma_\mu(q_0) U.
\end{equation}

The 2nd, 3rd, and 4th terms of Eq.~(\ref{M_g}) already contain
the first power of $r$ in the numerators,
thus here we may safely ignore
the difference between $p'$ and $p'_0$, $U'$ and $U'_0$, etc.

Then we proceed to determining the effective TPE amplitudes $F_i$
and computing corrections to observables,
as described in the previous section.

\section{Results and discussion}

\begin{figure}[ht]
  \hfill
  \parbox[t]{0.48\textwidth}
  {  \includegraphics[width=0.48\textwidth]{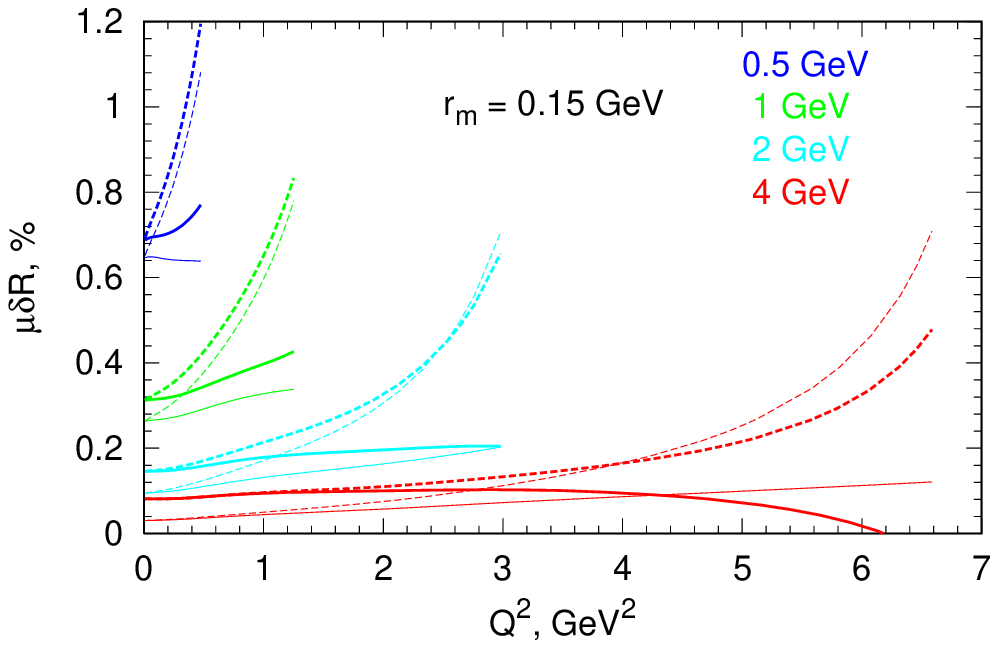}
     \caption{Bremsstrahlung correction to $\mu G_E/G_M$ ratio vs. $Q^2$
     at different beam energies, as labelled on the plot.
     Solid --- missing energy cut-off, dashed --- mising mass cut-off;
     thick --- full radiation, thin --- electron only.}\label{Q2plot}
  }
  \hfill
  \parbox[t]{0.48\textwidth}
  {  \includegraphics[width=0.48\textwidth]{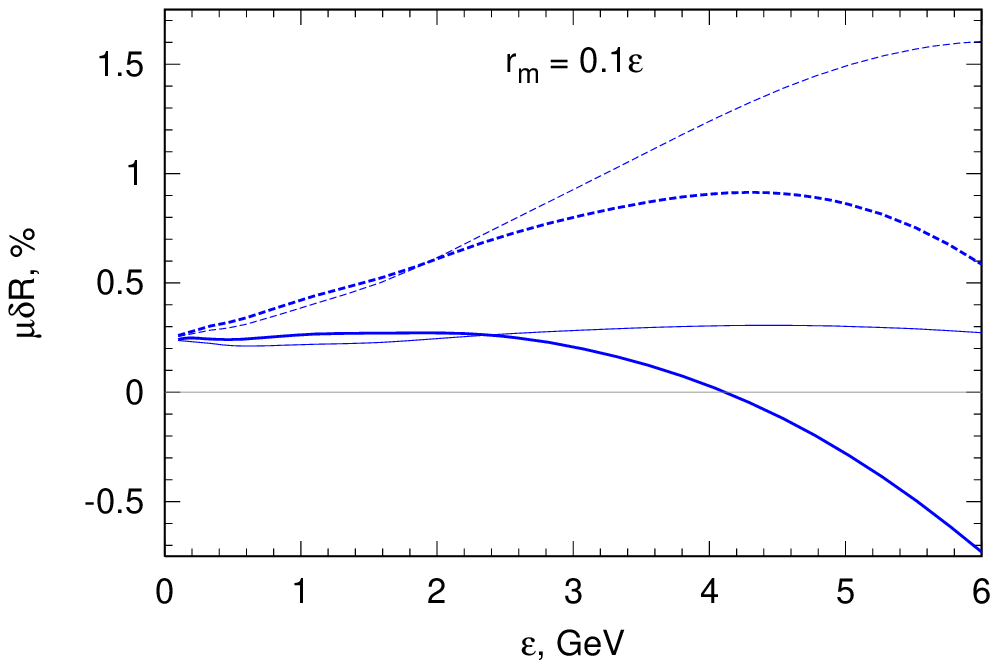}
      \caption{Bremsstrahlung correction to $\mu G_E/G_M$ ratio vs. beam energy
      at fixed scattering angle $90^o$.
      Curve types are the same as in Fig.~\ref{Q2plot}.
      }\label{Eplot}
  }
  \hfill
\end{figure}
\begin{figure}[t]
 \hfill
 \includegraphics[width=0.32\textwidth]{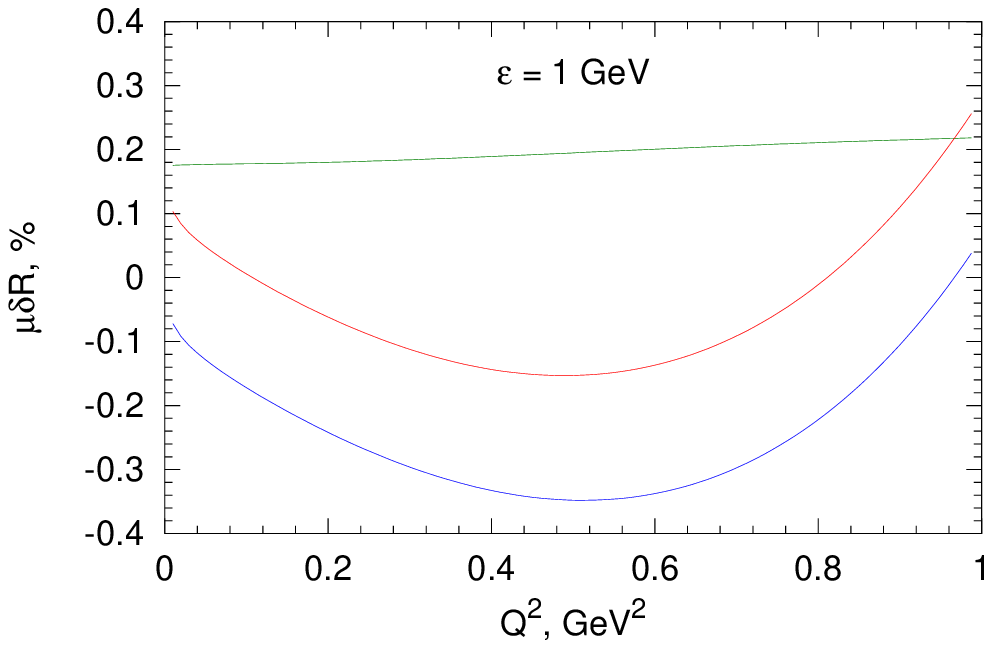}
 \hfill
 \includegraphics[width=0.32\textwidth]{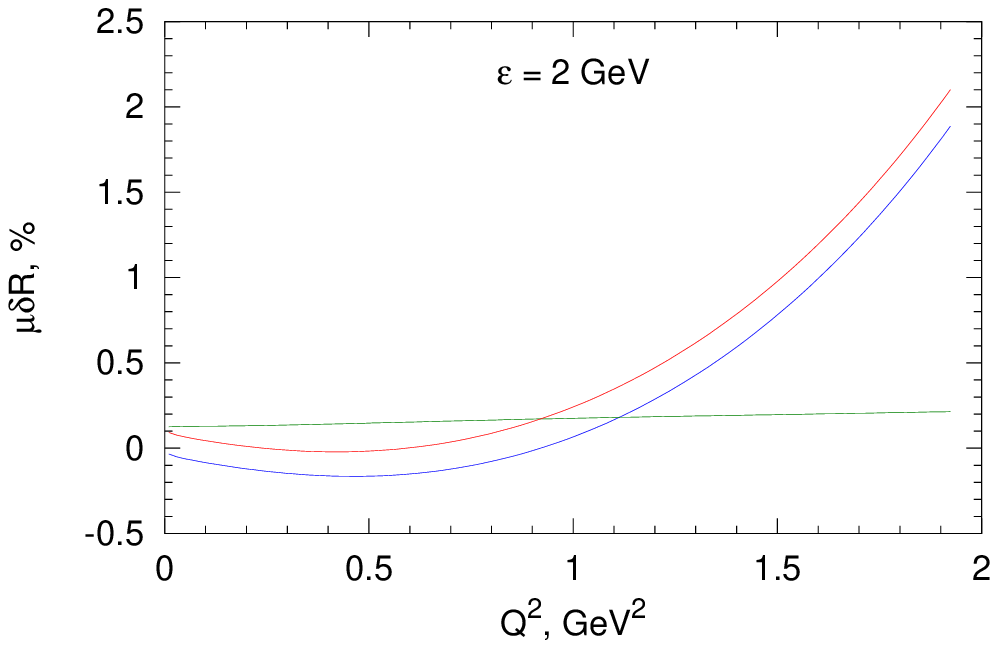}
 \hfill
 \includegraphics[width=0.32\textwidth]{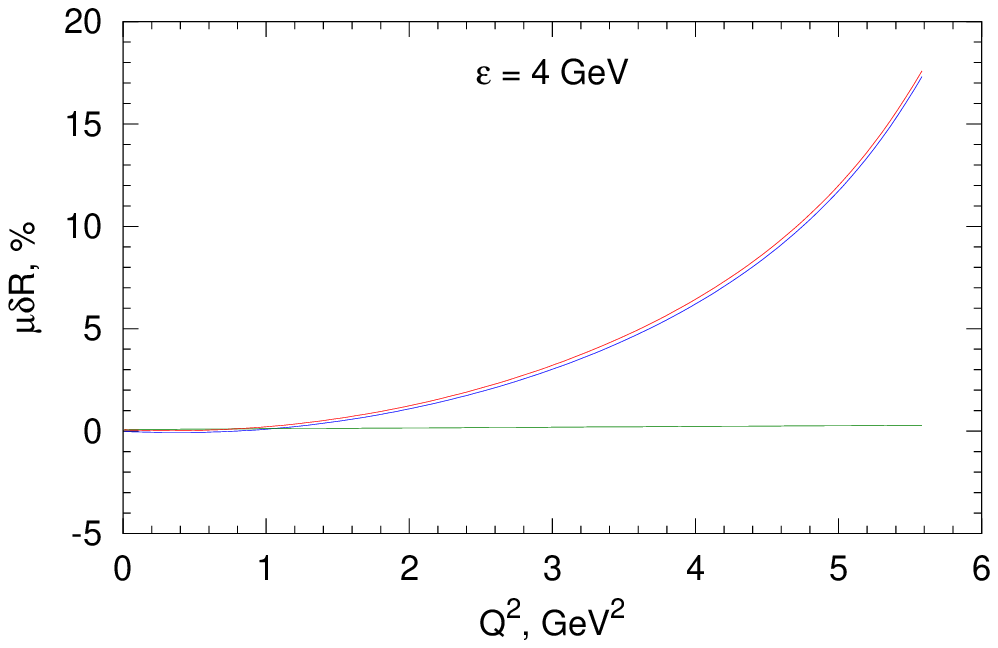}
 \hfill
 \caption{Radiative corrections to $\mu G_E/G_M$ ratio vs. $Q^2$,
   bremsstrahlung (green), TPE(blue), and total (red).
   Missing energy cut-off $r_m = 0.1\eps$.}
 \label{TOTplot}
\end{figure}

In all numerical calculations we use proton FF
parameterization by Arrington {\it et al.} \cite{ArringtonFF}.
Everywhere below $\eps$ is initial electron energy
(not to be confused with virtual photon polarization parameter).

Figure~\ref{Q2plot} displays the BS correction to $G_E/G_M$ ratio,
as measured via polarization transfer [Eq.~(\ref{R})],
at four different beam energies.
The missing energy cut-off is $r_m = 0.15\GeV$.
Since in our approximation the BS correction is proportional
to $r_m$, the transition to another $r_m$ value is straightforward.
The quantity shown in the figure is
$\mu\,\delta R = \delta R(Q^2)/R(Q^2=0)$.
It is more convenient to plot than the relative correction
$\delta R/R$,
since $R$ approaches zero at $Q^2 \sim 7\GeV^2$;
therefore the relative correction strongly grows,
even while $\delta R$ itself does not.
The dashed curves are obtained in the "missing mass" approach
with the cut-off $(p'+r)^2 - M^2 \le 2Mr_m = u_m$.
Thick curves results from the full calculation ($Z=1$),
thin ones --- including electron radiation only
(dropping two last diagrams in Fig.~\ref{diagr}
 or putting $Z=0$ in the r.h.s. of Eq.~(\ref{M_g})).

We see that the BS correction is typically quite small ($<1\%$),
and has a tendency to drop as energy increases.
This may indicate that the dimensionless expansion parameter
is really not $r_m/M$, but rather $r_m/Q$ or $r_m/E$.
The $Q^2$ dependence of the correction is weak, but at backward angles
(when $Q^2$ is close to its maximum) the missing mass approach
results in much larger correction.
We also see that the significant part of the full correction is
produced by proton radiation, especially at higher $\eps$ and $Q^2$.

The energy dependence of the BS correction at fixed lab. scattering angle
$90^\circ$ is shown in Fig.~\ref{Eplot}.
Here the missing energy cut-off is taken proportional
to the incident electron energy: $r_m = 0.1 \eps$.
The meaning of different curve types is the same as in Fig.~\ref{Q2plot}.
All four curves become close at $\eps \to 0$; this is clear,
since at $\eps \to 0$ the final proton remains
practically at rest ($p' \approx p$) and thus does not radiate.
At $\eps \gtrsim M$ full and "electron only" calculations
give very different results, as expected.

Comparing our results with the results of Refs.~\cite{Afa1,Afa2},
we note that the linearity of the BS correction
in $r_m$ at small $r_m \ll M$ is clearly seen in the various figures
from Refs.~\cite{Afa1,Afa2} (note that $u_m = 2M r_m$),
that is, the validity of the expansion in $r_m$
is supported by Refs.~\cite{Afa1,Afa2} as well.
{
At $\eps = 4\GeV$ the correction obtained in the ''missing mass'' approach,
has the same behaviour and magnitude
as shown in Fig.~4 of Ref.~\cite{Afa1}, but has opposite sign.
The origin of this discrepancy is unclear.
}

In Fig.~\ref{TOTplot} we plot the total radiative correction,
which is sum BS + TPE.
The TPE correction was calculated according to Refs.~\cite{ourDisp,ourP33}.
The BS is almost negligible with respect to TPE
at $\eps \ge 2 \GeV$; this is because the TPE correction
grows with the energy, contrary to BS.
At low energy the TPE correction is smaller and becomes
comparable to BS. With the cut-off $r_m = 0.1\eps$,
used to produce Fig.~\ref{TOTplot}, both corrections are negligible.
However, the magnitude of the BS correction (contrary to the TPE one)
substantially depends on the experimental details.
If different cut-off is used in an experiment, the correction
may become much larger (see e.g. Fig.~\ref{Q2plot}).
Thus we conclude that the BS corrections are of small importance
for prospective high-$Q^2$ experiments, but may be significant
and need to be more carefully analysed for low-$Q^2$ ones.

\section{Conclusions}

We have studied radiative corrections for the polarization transfer measurements of the proton FF ratio including both TPE and BS corrections.
The latter was calculated assuming both electron and proton can radiate.
Two approaches to the elastic event selection were considered: missing energy and missing mass cut-off.

Numerical calculation shows that:
\begin{enumerate}
\item The proton radiation yields a significant part of 
 the BS corrrection at $\eps \gtrsim M$ in both "missing energy" and "missing mass" approaches.
\item In the "missing mass" approach the correction strongly grows at
  large angles, whereas in the "missing energy" approach it does not.
\item The BS correction is small at high energies ($\eps \gtrsim M$), where the TPE correction is much larger.
  However there is no final reliable estimate of the TPE amplitude in this region; this is an important open problem.
  The significance of the BS correction at low energies depends on experimental details; thus it should be checked separately for each case.
\end{enumerate}

\appendix

\section{Angular integrals $\phi_{ab\mu}$}  \label{App:integrals}

The integral to calculate is
\begin{equation}
  \phi_{ab\mu} 
  = \frac{r_m}{16\pi^3} \int \frac{\rho_\mu d\Omega_{\rho}}{(k_a \rho)(k_b \rho)}
  = \frac{r_m}{16\pi^3} \bar\phi_{ab\mu}, 
\end{equation}
where $\rho = (1,\boldsymbol\rho)$ and ${\boldsymbol\rho}^2 = 1$.
Obviously
\begin{equation}
  \bar\phi_{ab\mu} = A_{ab} k_{a\mu} + A_{ba} k_{b\mu} + B_{ab} M e_{0\mu},
\end{equation}
where $e_0 = (1,\vec 0)$.
The coefficients $A$ and $B$ are easily expressed via
the scalar integrals
\begin{equation}
 \int \frac{d\Omega_\rho}{(k_a \rho)(k_b \rho)}
 = \frac{4\pi \ka_a \ka_b}{R} \ln \frac{k_a k_b + R} {m_a m_b},
 \qquad
 \int \frac{d\Omega_\rho}{k_a \rho}
 = \frac{4\pi}{\ka_a} \ln \frac{\eps_a + \ka_a}{m_a},
\end{equation}
where $\ka_a = |\vec k_a|$, $m_a = \sqrt{k_a^2}$
and $R = \sqrt{(k_a k_b)^2 - m_a^2 m_b^2}$.
The final result is
\begin{equation}
 \begin{split}
  &A_{ab}=\frac{4\pi}{\ka_a^2 \ka_b \sin^2\theta} \left[
        (\eps_a\ka_b - \eps_b\ka_a\cos\theta) \frac{1}{R} \ln\frac{k_a k_b+R}{m_a m_b} 
        - \ln\frac{\eps_b+\ka_b}{m_b} + \cos\theta\ln\frac{\eps_a+\ka_a}{m_a}
   \right], \\
  &B_{ab}=\frac{4\pi}{M\ka_a^2 \ka_b^2 \sin^2\theta} \left[
        - R \ln\frac{k_a k_b+R}{m_a m_b} 
        + (\eps_a\ka_b - \eps_b\ka_a\cos\theta) \ln\frac{\eps_b+\ka_b}{m_b}
        + (\eps_b\ka_a - \eps_a\ka_b\cos\theta) \ln\frac{\eps_a+\ka_a}{m_a}
   \right],
 \end{split}
\end{equation}
where $\eps_a = k_{a0}$ and $\theta$ is the angle
between $\vec k_a$ and $\vec k_b$.

There are three special cases. For $k_a = k_b$ we imply $\bar\phi_{aa\mu} = 2A_{aa} k_{a\mu} + B_{aa} M e_{0\mu}$, with
\begin{equation} 
 A_{aa} = \frac{2\pi}{\ka_a^3} \left[ \frac{\eps_a\ka_a}{m_a^2} - \ln\frac{\eps_a + \ka_a}{m_a} \right], \qquad
 B_{aa} = \frac{4\pi}{M\ka_a^3} \left[ -\ka_a + \eps_a \ln\frac{\eps_a + \ka_a}{m_a} \right]
\end{equation}
for $k_b = k_2 \equiv p = M e_0$
\begin{equation} 
 A_{2a} \equiv 0, \qquad
 A_{a2} = - \frac{4\pi}{M\ka_a^2} \left[ 1 - \frac{\eps_a}{\ka_a} \ln\frac{\eps_a + \ka_a}{m_a} \right], \qquad
 B_{a2} =   \frac{4\pi}{M^2\ka_a^2} \left[ \eps_a - \frac{m_a^2}{\ka_a} \ln\frac{\eps_a + \ka_a}{m_a} \right]
\end{equation}
and for $k_a = k_b = k_2$
\begin{equation} 
 A_{22} \equiv 0, \qquad B_{22} = \frac{4\pi}{M^3}.
\end{equation}

In Refs.~\cite{Afa1,Afa2}, authors consider a cut on the missing mass,
instead of the missing energy, as an event selection criterion.
This case is also covered by our approach.
All ingredients of the calculation remain unchanged
except the integrals $\phi_{ab\mu}$.
The integrals can be rewritten in fully covariant form as
\begin{equation}
  \phi_{ab\mu} = \frac{1}{(2\pi)^3}
    \int\limits_{pr \le M r_m} d^4 r \, \delta(r^2) \, \theta(r_0) \,
      \frac{r_\mu}{(k_a r)(k_b r)}.
\end{equation}
In the approach of Refs.~\cite{Afa1,Afa2},
the condition $pr \le M r_m$ is replaced by $(p+r')^2-M^2 = 2p'r \le u_m$,
thus the corresponding formulae can be obtained from the above
by substitution $p \leftrightarrow p'$, $r_m \to u_m/2M$.

\section{Effective TPE amplitudes} \label{App:Fi}

In this section we write down the effective TPE amplitudes $\delta F_i$,
which correspond to the BS amplitude as discussed in Sec.~\ref{Sec:BS}.
They are obtained as described in Sec.~\ref{Sec:BSdetail}.

Below $t\equiv q^2$, $F_m' = dF_m(q^2)/dq^2$, $F_2' = dF_2(q^2)/dq^2$,
$\kappa=F_2(0)$.
The quantities $A_{ab}$ and $B_{ab}$ are defined in the previous Appendix.
The electron mass $m$ is set to zero, except in front of $A_{11}$ and $A_{33}$,
which diverge as $1/m^2$ (see Eqs.(\ref{phi},\ref{mdiv})).
For brevity, we set $Z=1$ in the following equations.
The $Z$ dependence can be easily restored by putting
$A_{ab} \to z_a z_b A_{ab}$, $B_{ab} \to z_a z_b B_{ab}$,
with $z_1 = z_3 = 1$ and $z_2 = z_4 = Z$. 
Finally, $\delta F_m \equiv \delta F_1 + \delta F_2$, and $\delta X$ is the auxilliary quantity which enter the formulae for $\delta F_m$ and $\delta F_2$.

{
 \begin{multline}
\delta X = 
( (4 M^2-t)  t - (8 M^2-t) \nu  ) 
  ( ( t + \nu  )  A_{41} + ( t -  \nu  )  A_{43} )
 - 32 \nu M^4  A_{44} 
\\ + 4 t ( 4 M^2 -  t )
( 4 m^2  A_{11} + 2 t A_{13} - ( t + \nu  )  B_{13} )
\\ + t  ( 12 M^2 - 3 t -  \nu  )  
  ( ( t + \nu  )  A_{14} + ( t -  \nu  )  A_{12} )
\\  
  + t  ( 4 M^2 -  t + \nu  ) 
 (( t + \nu  ) A_{32} + ( t -  \nu  )  A_{34} )
\\ -  ( (4 M^2-t)  t + (8 M^2-3 t) \nu  ) 
   ( ( t + \nu  )  ( B_{14} + B_{32} )
    + ( t -  \nu  ) (B_{12}+B_{34}) )
\\ + 8 \nu  {( 2 M^2 - t ) }^2 B_{24} 
 + 8 \nu M^2 ( 2 M^2 -  t )   ( 2 A_{42} -  B_{22} -  B_{44} )
\end{multline}

\begin{multline}
\frac{\pi\delta F_m}{\alpha t r_m} =  
  \frac{1}{8}  [8 m^2 F_m (A_{11} - A_{33}) 
 + 4 t F_m (A_{13} - A_{31} - B_{13})
 -  ( F_2 - 2 F_m )   (( t + \nu ) ( A_{14} -  A_{32}  
  ) +  ( t -  \nu  )  ( A_{12} -  A_{34} )) 
\\ - 2 ( 2 M^2 ( F_2 - 2 F_m )  + ( t -  \nu  )  F_m )
   ( A_{43} - B_{12} + B_{34} ) 
\\ - 2 ( 2 M^2 ( F_2 - 2 F_m )  + ( t + \nu  )  F_m )
   ( A_{41} - B_{32} + B_{14} ) 
\\ + 2  t F_2 ( B_{14} + B_{34} )
 - 4 M^2 ( F_2 + F_m )  ( 2 A_{44} -  B_{22} ) 
\\ + 2 ( 4 M^2 F_m -  t ( F_2 + 2 F_m )  )  ( A_{42} -  B_{24} )
 + ( 4 M^2 F_1 + 2 t F_2 )  B_{44} 
 ] 
+ \frac{F_m'}{4 ( 4 M^2 - t + \nu  ) } \delta X
\end{multline}

\begin{multline}
\frac{\pi\delta F_2}{\alpha t r_m} = 
\frac{1}{2} [
  F_2 ( 2 m^2 (A_{11}-A_{33}) + t (A_{13} - A_{31} - B_{13}) )
 +  M^2  F_m
     ( A_{43} + A_{41} -  B_{12} -  B_{32} + B_{14} + B_{34} )
 ]
\\ + \frac{1}{4 ( 4 M^2 - t ) }
  [( t F_2 - 2 M^2 F_m )
   ( ( t + \nu  ) ( A_{32} -  A_{14} + B_{14})
    + ( t -  \nu  )  (A_{34} -  A_{12} + B_{34})) 
\\- (  ( 8 M^2 -  t )  F_2 - 2 M^2  F_m )
((t + \nu) ( A_{41} -  B_{32} ) + (t -  \nu)  ( A_{43} -  B_{12} ))
\\ + 2 ( 8 M^4 F_2 + t^2 F_2 - 2 M^2 t ( 5 F_2 -  F_m ) )
      ( A_{42} -  B_{24} )
\\ + 2 M^2 ( t F_2 - 4 M^2 ( 3 F_2 -  F_m ) )  ( 2 A_{44} -  B_{22})
 + 2 M^2 ( 4 M^2 F_1 + t ( 3 F_2 - 2 F_m ) )  B_{44}
 ]
\\-  \frac{\kappa}{4}  \nu  F_1  [B_{14} -  B_{34} ]
+ \frac{4 M^2 F_1 -  
   t F_2 + 2 ( 4 M^2 -  t )  t F_2'}{8 t 
  (4 M^2 - t )  ( 4 M^2 - t + \nu  ) } \delta X
\end{multline}

\begin{multline}
\frac{\pi\delta F_3}{\alpha t r_m} =  
  \frac{1}{4} [
  2 M^2 F_2 ( A_{43} + B_{32} - A_{41} - B_{12} + B_{14}-B_{34})
  -  \kappa( t  F_2 + 4 M^2  F_1)  ( B_{14}-B_{34} )
   ]
\end{multline}
\begin{multline}
\frac{\pi\delta F_5}{\alpha t r_m} = -\frac{\nu  F_m}{2 t (  
  4 M^2 - t + \nu  ) } [4 m^2 t A_{11} 
 + 2 t^2 A_{13}
 + t ( t + \nu  )  A_{14} 
 + t ( t - \nu  )  A_{12} 
\\ + 2 M^2 (( t + \nu  )  A_{41} + (t- \nu  )  A_{43}) 
 + 8 M^4 A_{44} - 2 {( 2 M^2 - t  
  ) }^2 B_{24} 
\\ + ( 2 M^2 - t )  ( 2 t B_{13}  
 + ( t + \nu  )  ( B_{14} + B_{32} )
 + ( t - \nu  )  (B_{12} + B_{34})
 - 2 M^2 ( 2 A_{42} - B_{22} - B_{44} )  )  ]
\end{multline}
\begin{multline}
\frac{\pi\delta F_8}{\alpha t r_m} = \frac{1}{16} [( t + \nu   
  )  F_2  (A_{14} + A_{32}) +  
  ( t - \nu  )  F_2 ( A_{12} + A_{34} ) 
\\ - 4 M^2 F_1 ( A_{41} - A_{43} - B_{12} + B_{32} )  -  
  2 ( 2 M^2 F_1 + t ( F_2 + \kappa  F_m )  )  
   ( B_{14} - B_{34} )  ]
\end{multline}
\begin{multline}
\shoveright{\delta F_4 = \frac{({\nu }^2 -t^2 )  F_2 -
  4 M^2 t  F_1 }{4 M^2 \nu  F_m}\delta F_5}
\end{multline}
\begin{multline}
\shoveright{\delta F_7 = \frac{t}{2\nu} \delta F_5}
\end{multline}

}

\section{Observables} \label{App:M2}

In the case of double-polarization experiment the amplitude squared
has the following general structure:
\begin{equation}
 |\bar \CM|^2 = \alpha^2
    \left( a + b_\mu s_\mu + c_\mu S_\mu + d_{\mu\nu} s_\mu S_\nu \right),
\end{equation}
where $s_\mu$ is incoming electron polarization and $S_\mu$
is the spin of the proton, either initial or final.
The quantities $a$, $b$, $c$ and $d$ are quadratic functions
of the generalized FFs $F_i$.
The coefficients $b$ and $c$ have the form
\begin{equation}
 b_\mu = \sum_{i,j} b_{ij\mu} \Im F_i F_j^*, \qquad
 c_\mu = \sum_{i,j} c_{ij\mu} \Im F_i F_j^*.
\end{equation}
and thus vanish in the Born approximation.
They give rise to so-called single spin asymmetries;
we do not need to consider them further.
On contrary, the expressions for $a$ and $d$ involve the $\Re$ sign
\begin{equation}
 a = \sum_{i,j} a_{ij} \Re F_i F_j^*, \qquad
 d_{\mu\nu} = \sum_{i,j} d_{ij\mu\nu} \Re F_i F_j^* 
\end{equation}
The corrections to double-polarization observables are related to
$O(\alpha)$ terms in $d_{\mu\nu}$.
The detailed expression for $d_{\mu\nu}$,
corresponding to the square of ampltude (\ref{FullAmpl}), is written below.
In the Born approximation there are only two non-zero FFs, $F_1$ and $F_2$.
Thus all terms, proportional to $F_i F_j^*$ with $i,j \ge 3$ are dropped
--- they are small as $O(\alpha^2)$. We have
\begin{eqnarray}
d_{\mu\nu} = -\frac{2}{mM} \Re \left\{
  4 m^2 M^2 q^2 g_{\mu\nu}^\perp F_m^2
+ 4 m^2 q^2 [P_\mu P_\nu - P^2 g_{\mu\nu}^\perp] F_m F_2
-   m^2 q^2 [4K_\mu P_\nu - \nu g_{\mu\nu}^\perp] F_m F_3 \right. \nonumber \\
-   M^2 q^2 [4P_\mu K_\nu - \nu g_{\mu\nu}^\perp] F_m F_4
+ 4 M^2 q^2 [K_\mu K_\nu - K^2 g_{\mu\nu}^\perp] F_m F_5
- e_\mu e_\nu F_2 F_5/4 \nonumber \\
+ 8M^2 F_7 \left[
   4m^2 F_m K_\mu P_\nu - (\nu K_\mu + q^2 P_\mu)
   \left( F_e K_\nu + \tfrac{\nu F_2}{4M^2}P_\nu \right)
 \right] \nonumber \\ \left.
+ 8m^2 F_8 \left[
   q^2 F_m P_\mu K_\nu + (\nu F_1 P_\mu - 4M^2 F_e K_\mu) P_\nu
 \right] \right\},
\end{eqnarray}
%
where $g_{\mu\nu}^\perp = g_{\mu\nu} - q_\mu q_\nu / q^2$,
$e_\mu = 4 \ve^{\mu\nu\sigma\tau} P_\nu K_\sigma q_\tau$.
Contracting $d_{\mu\nu}$ with $s_\mu = k_\mu/m - m p_\mu/pk$,
which corresponds to the longitudinally polarized electron, we obtain the
final proton polarization
\begin{equation}
 S_\mu \sim A K_\mu + B P_\mu,
\end{equation}
with
\begin{eqnarray}
A & = & 4 M^2 F_m^2
    - (4M^2\!-\!q^2) F_m \left( F_2 - \frac{4M^2}{\nu\!-\!q^2} F_4 \right) + \nonumber \\
 && + \nu F_m \left( F_3 + 2 F_8 - \frac{4M^2}{\nu\!-\!q^2} F_5 \right) 
    + 8 M^2 F_e F_7 \left( 1 + \frac{4M^2}{\nu\!-\!q^2} \right) \\
B & = & \nu F_m \left( F_2 - \frac{4M^2}{\nu\!-\!q^2} F_4 \right)
    + q^2 F_m  \left( F_3 + 2 F_8 - \frac{4M^2}{\nu\!-\!q^2} F_5 \right) - \nonumber \\
 && - \frac{2 F_7}{\nu\!-\!q^2} [ 4M^2 (\nu F_e - q^2 F_m) - \nu^2 F_2 ]
   + 2 F_1 F_8 ( 4M^2\!-\!q^2\!+\!\nu^2/q^2 ).
\end{eqnarray}
From this it is easy to obtain Eq.~(\ref{deltaR}).

\end{document}